\documentclass[aps,twocolumn,pra,showpacs,amsmath,nofootinbib]{revtex4}
\usepackage{graphicx}% Include figure files
\usepackage{dcolumn}% Align table columns on decimal point
\usepackage{bm}% bold math
\usepackage{epstopdf}
\usepackage{multirow}
\newcommand{\beq}{\begin{equation}}
\newcommand{\eeq}{\end{equation}}
\newcommand{\bea}{\begin{eqnarray}}
\newcommand{\eea}{\end{eqnarray}}

\begin{document}

\author{O. Achakovskiy}
\affiliation{Institute for Physics and Power Engineering, 249033 Obninsk, Russia}
\author{A. Avdeenkov}
\affiliation{Institute for Physics and Power Engineering, 249033 Obninsk, Russia}
\author{S. Goriely}
\affiliation{Institut d'Astronomie et d'Astrophysique,
ULB,  CP 226, B-1050 Brussels, Belgium}
\author{S. Kamerdzhiev\footnote{kaev@obninsk.com}}
\affiliation{Institute for Nuclear Power Engineering NRNU MEPHI, 249040 Obninsk, Russia}
\author{S. Krewald}
\affiliation{Institut f\"ur Kernphysik, Forschungszentrum J\"ulich, D-52425 
J\"ulich, Germany}

\title{Impact of the phonon coupling on the photon strength function}


\begin{abstract}
  The pygmy dipole resonance  and photon strength function in stable and unstable Ni and  Sn isotopes are  calculated 
%,?  and the PDR in 72Ni is predicted
within the microscopic self-consistent version of the extended theory of finite fermi systems which includes 
phonon coupling effects.
 The  Skyrme force SLy4 is used. A pygmy dipole resonance in $^{72}$Ni is predicted at the mean energy of 12.4 MeV exhausting 25.7\% of the total energy-weighted sum rule.
  The microscopically obtained  photon E1 strength functions are compared with available experimental data and used to calculate nuclear reaction properties. Average radiative widths and radiative neutron capture cross sections have been calculated taking the phonon coupling into account as well as the uncertainties caused by 
various microscopic level density models. In all three  quantities considered, the contribution of phonon coupling turned out to be significant and is found necessary to explain  available experimental data. 

\end{abstract}
\maketitle

The photon strength function (PSF) is a quantity of fundamental importance in the description of nuclear reactions involving electromagnetic excitations or de-excitations. In particular, the PSF is known to significantly affect the radiative neutron capture cross section for incident neutrons in the keV region, a range of energies of particular relevance in astrophysical  \cite{gor02,gor04} as well as nuclear engineering \cite{muhab} applications. 
Traditionally, the de-excitation PSF has been parametrized phenomenologically on the basis of simple Lorentzian-type functions \cite{brink55,ripl2,ripl3} and associated with the photoabsorption strength function on the basis of the Brink hypothesis \cite{brink55} which assumes that on each excited state, it is possible to build a giant dipole resonance (GDR) equivalent to the one observed in the reverse photoabsorption process. 
For the last decades, an important effort has been devoted to the determination of the low-lying strength, i.e. the tail of the GDR. This low-lying strength, in particular below the neutron threshold, is known to be of relevance in radiative neutron cross section. The presence of any extra strength with respect to the tail of the GDR, this so-called pygmy dipole resonance (PDR), has been found to exhaust typically about 1-2\% of the Energy Weighted Sum Rule (EWSR) but also to affect significantly the radiative neutron capture cross section and potentially the nucleosynthesis of neutron-rich nuclei by the r-process \cite{arnould07}. Recent experiments \cite{savran2013,toft,uts2011,tsoneva} have provided additional information about the PDR  and PSF structure that clearly cannot be explained by standard phenomenological approaches. It was also shown that in neutron-rich nuclei, the strength could be significantly more important than in nuclei close to the valley of $\beta$-stability \cite{gor02,gor04}. For all these reasons, there has been a growing interest in the investigation of the PDR energy region both by the low-energy nuclear physics community \cite{savran2013,paar2007}  and within the field of nuclear data \cite{gor02,gor04,ripl2,ripl3}. 

Since the presence,  strength and structures of a PDR cannot be predicted within the Lorentzian-type approach, microscopic models need to be applied. Mean-field approaches using effective nucleon interactions, such as the Hartree-Fock Bogoliubov (HFB) method and the quasi-particle random-phase approximation (QRPA)  allow for systematic self-consistent studies of the E1 strength functions. Provided damping and deformation corrections are included phenomenologically on top of the HFB+QRPA strength, such methods have proven their capacity to reproduce fairly well the photoabsorption data in the vicinity of the GDR.   However, as discussed below and as confirmed by recent experiments (see for example \cite{uts2011}), the  HFB+QRPA  approach fails to reproduce fine structures and needs to be renomalized on GDR data. To be exact, it should be complemented by the effect describing the interaction between the single-particle and low-lying collective phonon degrees of freedom, known as the  phonon coupling (PC) \cite{revKST,ave2011,kaev2014,vg}. Such an interaction was originally considered in the 
%so-called 
quasiparticle-phonon model \cite{Sol76,vg}.
 
 In this work, to go beyond the HFB+QRPA method, we use  the self-consistent version of the extended theory of finite fermi systems (ETFFS) \cite{revKST} in the quasi-particle time blocking approximation (QTBA) \cite{tselyaev}.  
Our ETFFS(QTBA) method, or simply QTBA,  includes self-consistently 
the QRPA and PC effects and the single-particle continuum in a discrete form. Details of the method can be found in Ref.~\cite{ave2011}. The method also allows us  to investigate the impact of the PC on nuclear reactions in both stable and unstable  nuclei.  To do so, we calculate  the microscopic PSFs in different Sn and Ni  isotopes and use them to estimate the impact of the PC on radiative neutron capture cross section as well as on the average radiative widths on the basis of modern nuclear reaction codes like EMPIRE \cite{empire} and TALYS \cite{koning12}. 

\vspace{0.2cm}

The strength function $S(\omega) = dB(E1)/dE$ \cite{revKST,ave2011}, related to  the PSF $f(E1)$  by  
$f(E1,\omega)[{\rm MeV}^{-3}] = 3.487\cdot10^{-7} S(\omega)[{\rm fm}^2{\rm MeV}^{-1}]$,
is calculated by the QTBA method  \cite{revKST,tselyaev} on the basis of the well-known SLy4 Skyrme force \cite{chabanat}.  
The ground state is calculated within the  HFB method using the spherical code HFBRAD~\cite{bennaceur}. 
The residual interaction for the (Q)RPA and  QTBA calculations  is derived as the second derivative of the Skyrme functional. In all our calculations
 we use  a smoothing parameter of 200 keV which effectively accounts for correlations beyond the considered PC.
   Such a choice guarantees a proper description of all three characteristics of giant resonances, including the width \cite{revKST}, and also corresponds to the experimental resolution of reference in the present work \cite{toft}.

\begin{figure}
\includegraphics[width=8.6cm,clip]{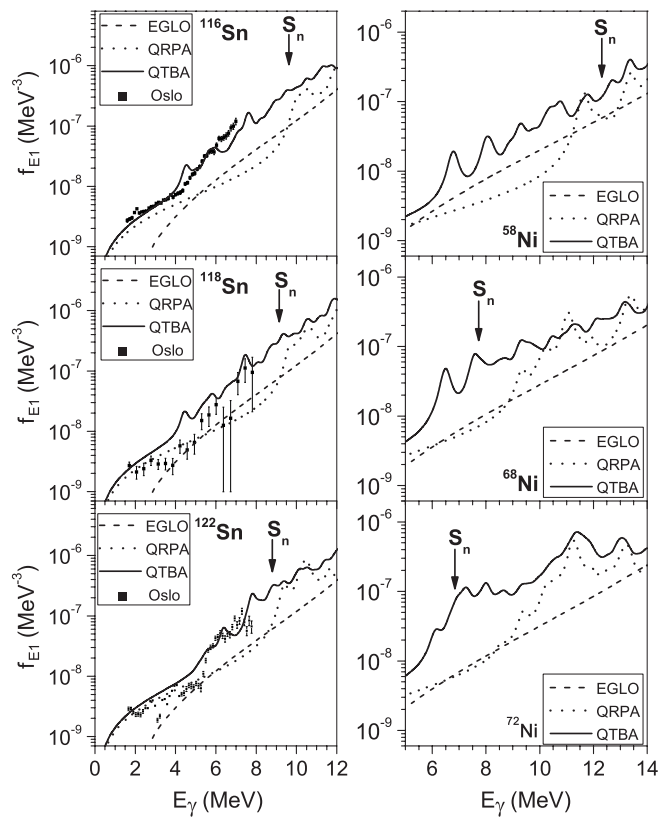}
\caption{ E1 PSF for $^{116,118,122}$Sn and for $^{58,68,72}$Ni in the low-energy PDR energy region. Dotted lines correspond to the self-consistent QRPA, full lines to the QTBA (including PC), and dashed lines to the EGLO model \cite{ripl2}. For Sn isotopes, experimental data are taken from \cite{toft}.}
\label{fig-1}       % Give a unique label
\end{figure}

In Fig.~\ref{fig-1}, the E1 PSF for specific even-even  Sn and Ni  isotopes is compared with experimental data obtained with the Oslo method \cite{toft} for Sn isotopes as well as the phenomenological Enhanced Generalized Lorentzian (EGLO) model \cite{ripl2}. 
It can be seen that {\it i)} in  contrast  to phenomenological models,  the structure patterns caused by both the QRPA and PC effects are pronounced in both Sn and Ni isotopes. Physically, the PC structures are caused by the PC poles at  $\omega = \epsilon_1 -\epsilon_2 - \omega_s $ or $\omega = E_1 + E_2 -\omega_s$, where $ \epsilon_1, E_1, \omega_s $  are single-particle, quasi-particle and phonon energies, respectively. Such a PC effect is seen to become significant above 3~MeV and below typically 9--10~MeV;
% which can be seen only  beginning from (9-10)MeV, 
{\it ii)} for   $^{118}$Sn and  $^{122}$Sn  isotopes, a reasonable agreement with experiment is obtained  within the QRPA below typically 5 MeV. 
For all three Sn isotopes, at E$>$5~MeV, the inclusion of PC effects is needed to reconcile predictions with experiment \cite{toft}; {\it iii)} globally, the EGLO description of the experimental data is noticeably worse than the one achieved by the QTBA.

In Table \ref{tab-1}, the integral parameters (mean energy $E$ and fraction of EWSR exhausted)  of the PDR are given for  three Ni isotopes, as predicted by both QRPA and QTBA (QRPA+PC) models.   
To compare results  in these three  nuclei,  a 6 MeV energy interval, which corresponds to the one where the PDR was observed in $^{68}$Ni,  is considered. In this interval, the PDR characteristics have been approximated, as usually, with a Lorentz curve by fitting the three moments of the theoretical curves  \cite{revKST}.
For $^{68}$Ni, a good agreement is obtained with experimental data of $E  \simeq 11$~MeV and about 5\% of the total EWSR  \cite{wiel}. 
A similar calculation was performed for $^{68}$Ni  \cite{lrt2010} using the relativistic QTBA, with  two phonon contributions additionally taking into account.
  For the PDR characteristics in $^{72}$Ni in the (8-14) MeV range, we obtain a mean energy $E=12.4$~MeV,  width $\Gamma = 3.5$~MeV and a large strength of 25.7\% of the EWSR. It should be noted that the main contribution to the $^{72}$Ni PDR is found in the (10-14) MeV interval which exhausts 13.9\% of the EWSR for QRPA and 23.2\% for QTBA. In this interval, two maxima can be observed (Fig~\ref{fig-1}). For this reason, the strength in the (10-14) MeV dominates and is globally equivalent to the one in (8-14) MeV.  A large PC contribution to the PDR strength is found in all isotopes (Table~\ref{tab-1}).    
     
\begin{table}[ht]
\centering
\caption{Integral characteristics of the PDR (centroid energy $E$ in MeV and fraction of the EWSR) in Ni isotopes 
calculated in the (8-14)~MeV interval for $^{58}$Ni, $^{72}$Ni and (7-13)~MeV interval for $^{68}$Ni (see text for details).}
\label{tab-1}
\begin {tabular}{ l l l l l l}
\hline
\hline
\multirow{2}{*}{Nuclei}&\multicolumn{2}{l}{QRPA}&\multicolumn{2}{l}{QTBA}\\
\cline{2-5}
&$E$&\%&$E$&\%\\
\hline
$^{58}$Ni&13.3&6.0&14.0&11.7\\
%\hline
$^{68}$Ni&11.0&4.9&10.8&8.7\\
%\hline
$^{72}$Ni&12.4&14.7&12.4&25.7\\
\hline
\hline
\end{tabular}
 \end{table}

\begin{figure}
\includegraphics[width=8.6cm,clip]{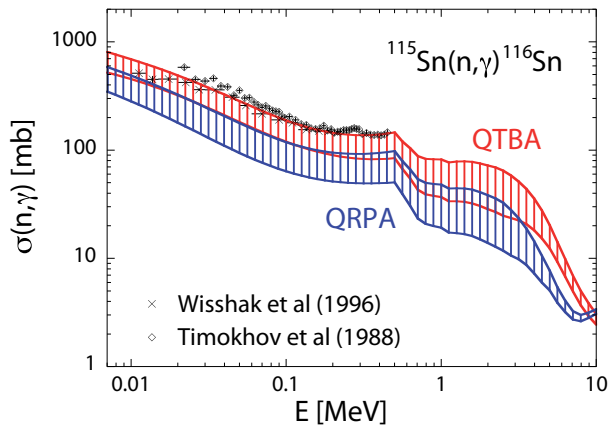}
\caption{(Color online) $^{115}$Sn(n,$\gamma$)$^{116}$Sn cross section calculated with the  QRPA (blue) and QTBA (red) PSF. The uncertainty bands depict the uncertainties affecting the NLD predictions \cite{ripl2,kon08,gor08,hil12}. Experimental cross sections are taken from Refs.~\cite{wisshak,timokhov}.}
\label{fig-2}   
\end{figure}

\begin{figure}
\includegraphics[width=8.6cm,clip]{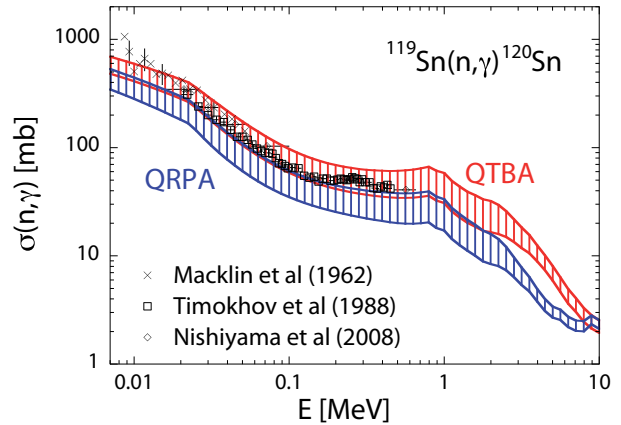}
\caption{ (Color online) Same as Fig.~\ref{fig-2} for $^{119}$Sn(n,$\gamma$)$^{120}$Sn.  Experimental  cross  section are taken from Refs.~\cite{timokhov,Macklin,Nishi} }
\label{fig-3}  
\end{figure} 

In Figs.~\ref{fig-2} and \ref{fig-3} we  present 
 the radiative neutron capture cross sections  estimated with the Hauser-Feshbach reaction code TALYS \cite{koning12} on the basis of the newly determined gamma-strength function. Similar results are obtained if use is made of the EMPIRE reaction code \cite{empire}. The calculations were  performed with different nuclear level density (NLD) models, including the back-shifted Fermi gas model \cite{kon08}, the Generalized Superfluid model (GSM) \cite{ripl2} and the HFB plus Combinatorial  model \cite{gor08,hil12}. The NLD is constrained by experimental neutron spacings and low-lying states, whenever available \cite{ripl3}.  As seen in Figs.~\ref{fig-2}-\ref{fig-3}, the agreement with experiment is only possible  when the  PC is taken into account. QRPA approach clearly underestimates the strength at low energies. This deficiency is often cured by empirically shifting the QRPA strength to lower energies or broadening the distribution \cite{gor02,gor04}.

To test the low-lying strength predicted within the various existing models, we also consider the average radiative widths of neutron resonances 
$\Gamma_{\gamma}$, known to be a property of importance in the description of the $\gamma$-decay from high-energy nuclear states. This quantity is used in nuclear reaction calculations, in particular, to normalize the PSF around the neutron threshold and is defined by \cite{belanova}
%the $\gamma$-ray strength.
\begin{equation}
 \Gamma_{\gamma} = \sum_{I=|J-1|}^{J+1}\int_{0}^{S_n}\epsilon^3_{\gamma}f_{E1}
 (\epsilon_{\gamma})\dfrac{\rho(S_n-\epsilon_{\gamma},I)}{\rho(S_n,J)}d\epsilon_{\gamma},
\end{equation}
where $\rho$ is the NLD and $J$ the spin of the initial state in the compound nucleus.
Extended compilation of experimental data  for $\Gamma_{\gamma}$ can be found in Refs.~\cite{ripl2,ripl3,muhab}. We have calculated the $\Gamma_{\gamma}$ values for 13 Sn and Ni isotopes on the basis of the EMPIRE code \cite{empire} for the 3 different PSF models, namely EGLO, our SLy4+QRPA and the present QTBA, together with different NLD prescriptions, namely the GSM \cite{ripl2} and the microscopic HFB plus combinatorial model \cite{gor08}. The predictions are compared in Table \ref{tab-2} with experimental data~\cite{muhab}, whenever available, and with existing systematics \cite{ripl2,ripl3}.  
%As far as we know, these are the first calculations of the  $\Gamma_{\gamma}$ values performed with PC. 
As seen in Table.~\ref{tab-2},
   the  PC effect in stable nuclei significantly increases  the QRPA contribution and improves the agreement with the  systematics. Except for $^{122}$Sn and  $^{124}$Sn, where the increase is limited, the PC leads to an enhancement of  about  50 to 200\%. 
   
   \begin{table*}[ht]
\centering
\caption{Average radiative widths $\Gamma_{\gamma}$ (meV) for s-wave neutrons. For each approach (EGLO, QRPA and QTBA) two NLD models are considered: the phenomenological GSM \cite{ripl2} (first line) and the microscopic HFB plus combinatorial model \cite{gor08} (second line). See text for details.}
\label{tab-2}
\begin {tabular}{ l l l l l l l l l l l l l l}
\hline
\hline
&$^{110}$Sn&$^{112}$Sn&$^{116}$Sn&$^{118}$Sn&$^{120}$Sn&$^{122}$Sn&$^{124}$Sn&$^{132}$Sn&$^{136}$Sn&$^{58}$Ni&$^{62}$Ni&$^{68}$Ni&$^{72}$Ni\\
\hline
\multirow{2}{*}{EGLO}&147.4&105.5&72.9&46.6&55.0&56.6&49.9&398&11.1&1096&794&166&320\\
&207.9&160.3&108.9&106.7&124.3&110.2&128.7&4444&295.0&2017&1841&982.2&86.4\\
%\hline
\multirow{2}{*}{QRPA}&45.6&34.4&30.4&22.1&23.8&27.9&22.3&133&11.2&358&623&754&83.8\\
                     &71.0&49.7&44.3&40.3&43.0&50.1&68.9&4279&447.8&450.8&490.9&406.4&46.7\\
%\hline
\multirow{2}{*}{QTBA}&93.5&65.7&46.8&33.1&34.1&35.8&27.9&148&12.3&1141&1370&392&154\\
                     &119.9&87.0&58.4&58.1&61.5&64.0&84.8&4259&509.2&1264&2117&2330&53.8\\
%\hline
\multirow{2}{*}{Exp.} \cite{muhab}&&&&117 (20)&100 (16)&&&&&&2000 (300)&&\\
$\qquad$ \cite{ripl2}&&&&80 (20)&&&&&&&2200 (700)&&\\
%\hline
System.&112&109&107&106&105&104&103&85&73&2650&1300&420&320\\
\hline
\hline
\end{tabular}
 \end{table*}

 Our  $\Gamma_{\gamma}$  results for $^{118}$Sn, $^{120}$Sn and $^{62}$Ni, for which  experimental data (not systematics) exists,  are of special  interest.
On the basis of the QTBA strength and the microscopic HFB plus combinatorial NLD \cite{gor08}, we obtain a good  agreement with experiment for $^{62}$Ni, and reasonable for $^{118}$Sn and $^{120}$Sn. Note that on top of the E1 strength, an M1 contribution following the recommendation of Ref.~\cite{ripl3} is included in the calculation of $\Gamma_{\gamma}$. 
%{\bf (TO BE CONFIRMED !?)}.
%This fact  probably means that the M1 contribution to $\Gamma_{\gamma}$ is not large
The M1 resonance contribution to $\Gamma_{\gamma}$ has been estimated using the GSM NLD model and the standard Lorentzian parametrization \cite{ripl3} with a width $\Gamma = 4$~MeV (note that such a large $\Gamma$ value is open to question, as discussed in Ref.~\cite{kaevkov2006}). Such a contribution is found to be of the order of (10-12)\% of the values in the first line of Table.~\ref{tab-2} for Sn isotopes and 4\%,3\%, 22\% and 16\%  for $^{58}$Ni, $^{62}$Ni, $^{68}$Ni and $^{72}$Ni, respectively.
The agreement  of the $\Gamma_{\gamma}$ values with experiment is found to deteriorate if use is made of the EGLO or QRPA strengths, but also of the GSM NLD. One can also see  that for stable nuclei, the combinatorial NLD model results  are  in a better agreement with the systematics \cite{ripl2} than those obtained with the GSM model.  As far as the EGLO model is concerned, we see that similar conclusions can be drawn.

In conclusion, the microscopic  E1 PSFs for 13 Sn and Ni isotopes have been calculated  within the self-consistent QTBA approach which takes into account the QRPA and PC effects and uses the known SLy4 Skyrme force. They have been  used to calculate the radiative neutron capture cross sections and  average radiative widths of neutron  resonances. A reasonable agreement with available experimental data has been obtained thanks to the PC which turns out to contribute significantly. 
  
Our results show  the necessity to include the PC effects into the  theory of radiative nuclear data for low-energy nuclear physics  for both stable and unstable nuclei. The QTBA method remains to be applied to the bulk of nuclei of astrophysical interest, but also to be compared with alternative approaches which  take the PC into account.

The work has been supported within  the   cooperation between INPE NRNU MIPHI  and Institute f\"ur Kernphysik FZ J\"ulich.
We  acknowledge E. Abrosimova, expert for International Research Funding and Cooperation of the Peter Gr\"unberg Institute/ J\"ulich Centre for Neutron Science, Forschungszentrum J\"ulich 
for significant  help within this project.  S. Ka-ev acknowlegdes useful discussions with Dr. V. Furman.

\end{document}